\documentclass[twocolumn]{revtex4}
\usepackage{graphicx,epsf}

\begin{document}


\title{ A Multigrid Algorithm for Sampling Imaginary-Time Paths in Quantum 
Monte Carlo Simulations }

\author{C.H. Mak\thanks{Author to whom correspondences should be addressed.} 
  and Sergey Zakharov\\
  Department of Chemistry, 
  University of Southern California, \\
  Los Angeles, California 90089-0482, USA }

\date{\today}



\begin{abstract} 

We describe a novel simulation method that eliminates the slowing-down problem
in the Monte Carlo simulations of imaginary-time path integrals near the
continuum limit.  This method combines a stochastic blocking procedure with
the multigrid method to rapidly accelerate the sampling of paths in a quantum
Monte Carlo simulation, making its dynamics more ergodic.  The effectiveness
and efficiency of this method are demonstrated for several one-dimensional 
quantum systems and compared to other standard and accelerated methods.

\end{abstract}

\maketitle


\section{Introduction}

The path integral formalism \cite{65feya,72feya} can be used to describe 
the statistical mechanics of a quantum system.  In the path integral 
representation, every quantum particle maps onto a cyclic gaussian string.  
In most path integral
simulations, the quantum strings are first discretized and then the sum over
all paths is carried out by Monte Carlo sampling.  In the 
discretized form, the path of a quantum particle is isomorphic to a 
classical gaussian ring polymer \cite{81cha4078}, and the statistical weight 
associated with each particle (in the canonical ensemble) takes the 
following form:
\begin{eqnarray} 
W &=& \exp \left( -\sum_{j=1}^{P} 
\frac{ \vert x_{j+1} - x_{j} \vert^2 }{2\lambda} - \epsilon V(x_{j}) \right), 
\label{eq:primative}
\end{eqnarray}
where $x_j$ is the position of the $j$-th bead on the ring, 
$\lambda = \epsilon\hbar^2/m$, $\epsilon = \beta/P = 
(Pk_BT)^{-1}$, $m$ is the mass, $T$ is the temperature and the bead indices 
$j$ are cyclic (i.e. $P+1 = 1$).  

To obtain the correct quantum limit, the continuum limit $P \to \infty$ 
must be taken.  In this limit, however, the harmonic bonds between 
successive beads on the ring polymer become very stiff.  This
causes the simulation to slow down dramatically on approach to the 
continuum limit, in a way that is very similar to a system undergoing 
a second-order phase transition.  As a result, a Monte Carlo algorithm 
employing only local updates will equilibrate extremely slowly.

This slowing-down problem in path-integral simulations can be remedied
in two ways.  First, one can use a more accurate approximation for the
short-time propagator in the path integral instead of the ``primitive'' 
approximation in Eqn.(\ref{eq:primative}), 
with the hope that not too many beads 
will suffice to accurately approximate the path integral \cite{95cep279}.
The second way is to devise Monte Carlo methods which employs nonlocal 
updates to hopefully remove or at least ameliorate the slowing-down problem.

In this paper, we are concerned with the second approach.  Several 
alternatives to the canonical single-particle Metropolis Monte Carlo method 
\cite{53met1087} have been proposed for this purpose.  First, there is
the so-called ``staging'' method 
\cite{84pol2555,85spr4234,85spr545,93tuc2796}.  
It attempts to reduce the correlation 
among the beads on the polymer by transforming to a new set of coordinates 
which diagonalizes the kinetic part of the action.  However, in the 
presence of a nonzero potential $V$, the transformed coordinates become 
correlated to each other through $V$ again.  To truncate these correlations, 
the staging method performs this transformation for one short segment of 
the ring at a time (hence the name ``staging'').  Updates in the transformed 
coordinates are done by direct sampling from independent distributions, 
but the new coordinates are accepted or rejected together based on a 
Metropolis criterion for the potential part of the action.
A second method, conceptually similar to the first one, is known as the 
Fourier path integral method \cite{86fre3536,86fre931,92lob4205,01mie621}.  
Here the Monte Carlo moves are performed
in the Fourier modes of the path, which also diagonalizes the kinetic part of 
the action.  But the Fourier modes are also correlated through $V$.  
A third method, which is also based on similar ideas as the first two, is 
the bisection method\cite{95cep279,98cha2123}.  Instead of generating 
the Fourier modes of the path, the bisection method samples the midpoint of
a large segment of a free-particle path and accept or reject it using an 
approximation for the long-time action that includes effects of the 
potential.  If the midpoint is accepted, the midpoints of the two shorter 
subsegments on each side of the midpoint are then generated, and this 
process is iterated until every bead on the entire segment is generated.
Finally, there is a multigrid-based method, first applied to path integrals
by Janke and Sauer \cite{93jan499,96jan488}.  They propose moving whole blocks
of neighboring beads on multiple length scales using a Metropolis algorithm, 
and they cycle through the different length scales in a systematic way.

In this paper, we describe another Monte Carlo method.  This method is 
actually related to the four methods described above, but as we will show, 
its formulation is more general, its applications are more powerful and 
its efficiency is much higher than the previous methods.  Our method 
combines a stochastic blocking procedure, often referred to as the 
Swendson-Wang method \cite{87swe86}, with multigrid ideas 
\cite{87bria,89goo2035} in an attempt
to formulate a set of equilibrium stochastic dynamics that is highly
ergodic for path integral simulations.  The idea 
for this type of multigrid method was first proposed 
by Kandel {\em et al.} for an Ising model at criticality \cite{88kan1591}.

The method will be described in Sections \ref{sect:general}--\ref{sect:PI}.  
Section \ref{sect:general} provides the general concept of the method, and 
Section \ref{sect:PI} applies the concept to 
the path integral problem.  
Section \ref{sect:results} will compare the method against others
for several examples of 1-dimensional systems with single- and double-well 
potentials.

\section{General Formulation} \label{sect:general}

The general idea of the multigrid Monte Carlo method has been described
in detail by Kandel {\em et al.} \cite{88kan1591} for the Ising model.  
We will not attempt to reproduce all the details here.  Instead, we will 
summarize the essentials in this section, using a language which is 
closer to path integrals.  The specific application of these ideas to 
path integral simulations will be described in detail in Sect.~\ref{sect:PI}.

Consider a system with partition function
\begin{eqnarray}
Q &=&  \int dx_1 \cdots dx_N e^{-{\cal S}}, \label{eq:Q}
\end{eqnarray}
where ${\cal S} = \sum_\alpha u_\alpha$, and 
each $u_\alpha$ is a real-valued interaction term involving any number 
of the $N$ particles in the system.  Of course, all classical systems, as 
well as quantum systems that can be mapped onto isomorphic classical 
polymeric systems through the path integral formalism, have partition 
functions of this form.  The correlations among the particles arise from
the interactions $u_\alpha$.

The multigrid Monte Carlo method is based on a combination of the stochastic 
blocking and multigrid ideas.  We will first describe the stochastic blocking 
procedure, often referred to as the ``unigrid'' method.  
To accelerate the dynamics of the 
system, the unigrid method proceeds in two stages.  First, with the current
configuration $X = \{x_1, \cdots x_N\}$, we attempt to 
remove some of the correlations from among the particles by ``killing'' 
the interaction terms $u_\alpha$ one by one:  For each $u_\alpha$, we 
consider either ``deleting'' it entirely from the action ${\cal S}$ 
with probability $p_d = c_\alpha \exp(u_\alpha)$ or ``freezing'' it with
probability $p_f = 1 - p_d$.  If an interaction is deleted, the ensuring
simulation can update $X \to X^\prime$ without any regard to $u_\alpha$.  
If on the other hand $u_\alpha$ is frozen, the ensuring simulation must 
not change the value of $u_\alpha$ during any update $X \to X^\prime$.  
To ensure that $p_d$ and $p_f \in [0,1]$, the coefficient 
$c_\alpha$ must be chosen to be smaller than $\exp(-u_\alpha^*)$, where
$u_\alpha^*$ is the largest possible value for $u_\alpha$.  
After all the interactions have been killed (deleted or frozen), the
particles can be divided into separate clusters -- particles in the same
cluster are connected by frozen bonds, while particles in different 
clusters are no longer correlated with each other.

In the second stage of the simulation, we can update each 
cluster separately with a Monte Carlo move that preserves the frozen bonds 
inside that cluster.  After all the clusters have been updated, we can
restore the interaction terms and repeat the procedure starting from the
first stage again, or we can use a few local Metropolis moves to update 
the system before starting the stochastic blocking procedure again.
It can easily be shown that this two-stage procedure satisfies detailed 
balance and therefore produces the correct statistical sampling\cite{87swe86}.

Under the unigrid method, interactions that are strong will
more likely be frozen and those that are weak will more likely be deleted.
This operation aims to remove some of the interactions from the system and this
can potentially make the subsequent updates more ergodic.  Whether this is
actually the case will depend on two factors: (1) whether the length scale
of the unconnected clusters are actually small enough, and (2) whether there 
exists an efficient way to update each cluster without disturbing the frozen 
interactions.  In reality, the length scale of unconnected clusters 
resulting from the stochastic blocking procedure can still be quite large, and 
hence a lot of the correlations remain in the system.  In addition, 
for systems with continuous coordinates, finding an efficient way to update 
all the particles in an unconnected cluster while preserving the frozen 
interactions is not always trivial.  Therefore, the stochastic blocking 
procedure ends up not being as useful as it may appear.

To repair this and to completely remove the residual correlations, we 
incorporate multigrid ideas and try 
to force the clusters to break up into smaller pieces of varying 
length scales.  To achieve this, the multigrid method first
divides the particles into 
sets, each having a different length scale.  In this context, the definition 
of the ``length scale'' should be based on an intuitive understanding of the
physical origin of the correlations in the system.  (For example, in 
path integrals, the dominant correlations among the beads on a ring 
originate from the harmonic bonds, so these correlations can be decomposed 
into a hierarchy of gaussian fluctuations on different length scales.)  
After a definition of these sets is made, we proceed as before but with 
the stochastic blocking procedure applied to {\em only} particles belonging
to a single length scale.  As such, we kill all the interactions between
particles of that length scale, while the other interactions 
among particles on all other length scales are kept
alive.  The result of the blocking procedure produces clusters that 
are unconnected by interactions on the current length scale, but these 
``unconnected'' clusters are not totally independent because they are 
still correlated with each other through the interactions that are kept 
alive.  We update the unconnected clusters as before, but to maintain 
detailed balance, each update will also have to be accepted or rejected 
using a Metropolis criterion based on all the live interactions that are linked 
to that cluster.  After updating all the ``unconnected'' clusters on one
length scale, we can proceed to another one.  In this manner, 
the multigrid method
systematically breaks up all the remaining correlations on every length
scale and the slowing-down problem can be completely eliminated.
It can also be shown that this multigrid procedure satisfies detailed
balance and therefore produces the correct statistical sampling\cite{88kan1591}.

\section{Application to Path Integral Simulations} \label{sect:PI}

In this section, we discuss the application of the method in 
Sect.~\ref{sect:general} to path integral simulations.  Here, we assume
a 1-dimensional particle in a potential $V$.  Generalization to higher
dimensions and many particles is straightforward.

\subsection{Definition of Length Scales} \label{sect:LS}

Before describing the unigrid and multigrid algorithms, we define the 
concept of length scales in a path integral simulation.  Let the number
of beads on the ring $P$ be equal to  $2^L$, where $L$ is a positive integer.
(Since the path is cyclic, bead $0$ is identical to bead $2^L$.)  
We divide the beads into different ``levels'' $\ell = 0, 1, \cdots L$, 
such that $\ell = \{ 1\times2^\ell, 3\times2^\ell, 5\times2^\ell, \cdots  \}$.
For examples, $\{1, 3, 5, \cdots\}$ would belong to $\ell=0$, $\{2, 6, 10, 
\cdots\}$ to $\ell=1$, $\{4, 12, 20, \cdots\}$ to $\ell=2$, etc.

Using these definitions, we say the full path is of length scale $L$.  
Starting with the full path, we divide it into segments of different length
scales at specific endpoints.  Bisecting the full path yields two equal-length
segments of length scale $L-1$, the first one having endpoints $0$ and 
$2^{L-1}$ and the second $2^{L-1}$ and $2^L$.  Bisecting each of these 
$L-1$ length scale 
segments again, we get four segments of length scale $L-2$, and so on.

In the absence of a potential $V$, every path segment of every length 
scale can be sampled independently from a gaussian distribution.  Therefore, 
even though beads on different levels $\ell$ are connected with each other via 
the kinetic energy springs, the path segments on all length scales can actually 
be generated in a completely uncorrelated manner.  But the presence of a 
nonzero potential $V$ introduces correlations back into the path segments, 
and they can no longer be sampled independently.   The potential produces
a ``confinement'' effect on the path, which couples path segments of certain 
length scales.  The length scale of these correlations depends on the spatial
extent of the confining potential as well as the temperature.   For a 
fixed temperature, a broad and shallow potential produces less correlation
than a steep narrow potential.  More complicated potentials (those frequently
present in condensed systems) may produce correlations on multiple length 
scales.  For example, a bistable potential produces two length scales, one for
intra-well quantum fluctuations and the other for inter-well fluctuations.
The goal of the stochastic blocking method is to remove some of these 
correlations produced by the potential, and the multigrid method furthermore 
refines it by attempting to remove these correlations over {\em all} length 
scales.

\subsection{Stochastic Blocking: The Unigrid Method} \label{sect:SB}

\begin{figure}
\includegraphics[width=0.8\columnwidth]{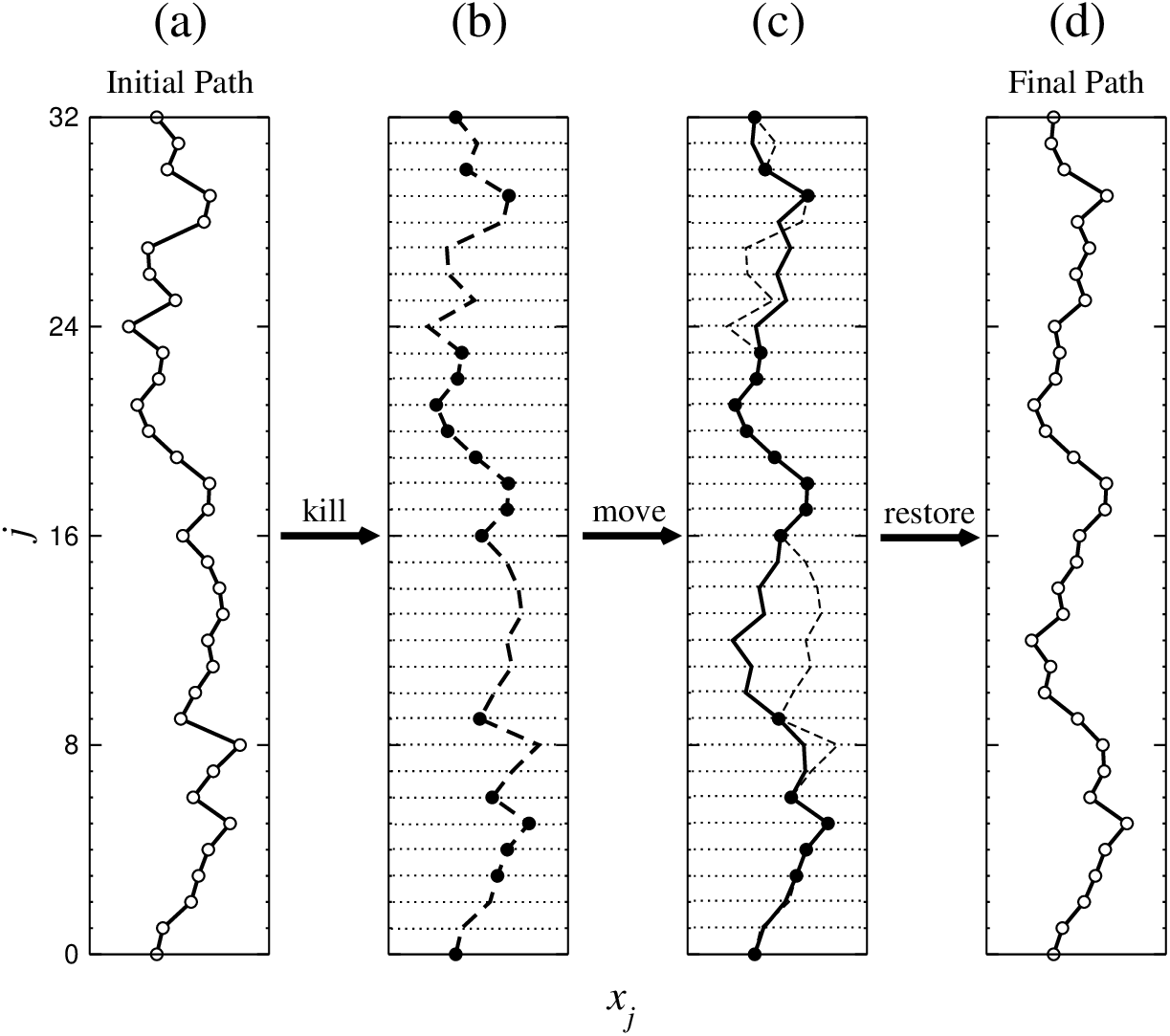}
\caption[]{
\label{fig:unigrid}
Illustration of the unigrid stochastic blocking procedure.  (See text
for details.)
}
\end{figure}

The stochastic blocking method aims at killing the correlations that 
come from the potential terms $V(x_j$) in Eqn.(\ref{eq:primative}).  
Following the ideas described in Sect.~\ref{sect:general}, we kill 
every $V(x_j)$ by either deleting it or freezing it.  This is illustrated
in Fig.~\ref{fig:unigrid} for a path with $L=5$.  Panel~(a) shows the
original path, with the $V(x_j)$ on all beads $i$ depicted pictorially as
open circles.  The stochastic blocking is done by first freezing 
the endpoints of the full path at $j = 0 = 2^L$ and then killing every 
potential term $V(x_j)$ for $0 < j < 2^L$.  This operation is represented in
panel~(b).  The dotted lines highlight the beads on which the killing 
operation is performed, and the result of each killing operation is either 
a frozen bead (represented by a closed circle) or a deleted bead 
(represented by the absence of a circle).   The path breaks up into frozen 
segments consisting of beads $0(=32)$, $3$--$6$, $9$, $16$--$23$ and 
$29$--$30$ and the intervening deleted segments.  The frozen segments 
can not be moved, but the deleted segments can be sampled independently 
from gaussian distributions for free-particle paths of various lengths.  
Panel~(c) shows the new path after the move (solid line) compared to the 
initial path (dashed line).  All the potential terms $V(x_j)$ are finally 
restored resulting in the final path in panel~(d).
This completes one pass and the next pass begins anew with the killing 
procedure performed on the path in panel~(d).
(Since the origin of the cyclic path at $j=0=2^L$ is always frozen in this 
method, we select a different origin at random on every pass to 
maintain ergodicity.)

Before moving on to the multigrid implementation, we want to mention one 
minor difference between what is described here and the original 
implementation of the 
stochastic blocking procedure due to Swendson and Wang \cite{87swe86} as 
summarized in Sect.~\ref{sect:general}.  The stochastic blocking procedure 
was originally applied to the Ising model, a system where the state of each 
particle belongs to a finite discrete set and every potential 
term $u_\alpha$ in the action of the Ising model is bounded from above by
$u_\alpha^*$.  As discussed in Sect.~\ref{sect:general}, to ensure that 
the deletion probabilities $p_d \in [0,1]$, $c_\alpha$ can be chosen to be 
less than $\exp(-u_\alpha^*)$ for every potential term.  In our path 
integral application, however, the coordinate of each bead is a continuous 
variable and the potential is in general {\em not} bounded from above.  
This means that to strictly satisfy the requirement that $p_d \leq 1$, 
$c_\alpha$ must be chosen to be $0$ which then results in no deletion at
all.  In practice, we can circumvent this minor problem by choosing a 
sufficiently large $u_\alpha^*$ and monitors the frequency at which the
bound $p_d \leq 1$ is violated during the simulation.  By adjusting 
$u_\alpha^*$ to yield a bound violation frequency of less than 0.01\%, 
we can maintain a relatively high deletion ratio while introducing negligible
errors to the results.

\subsection{The Multigrid Method} \label{sect:MG}

\begin{figure}
\includegraphics[width=0.8\columnwidth]{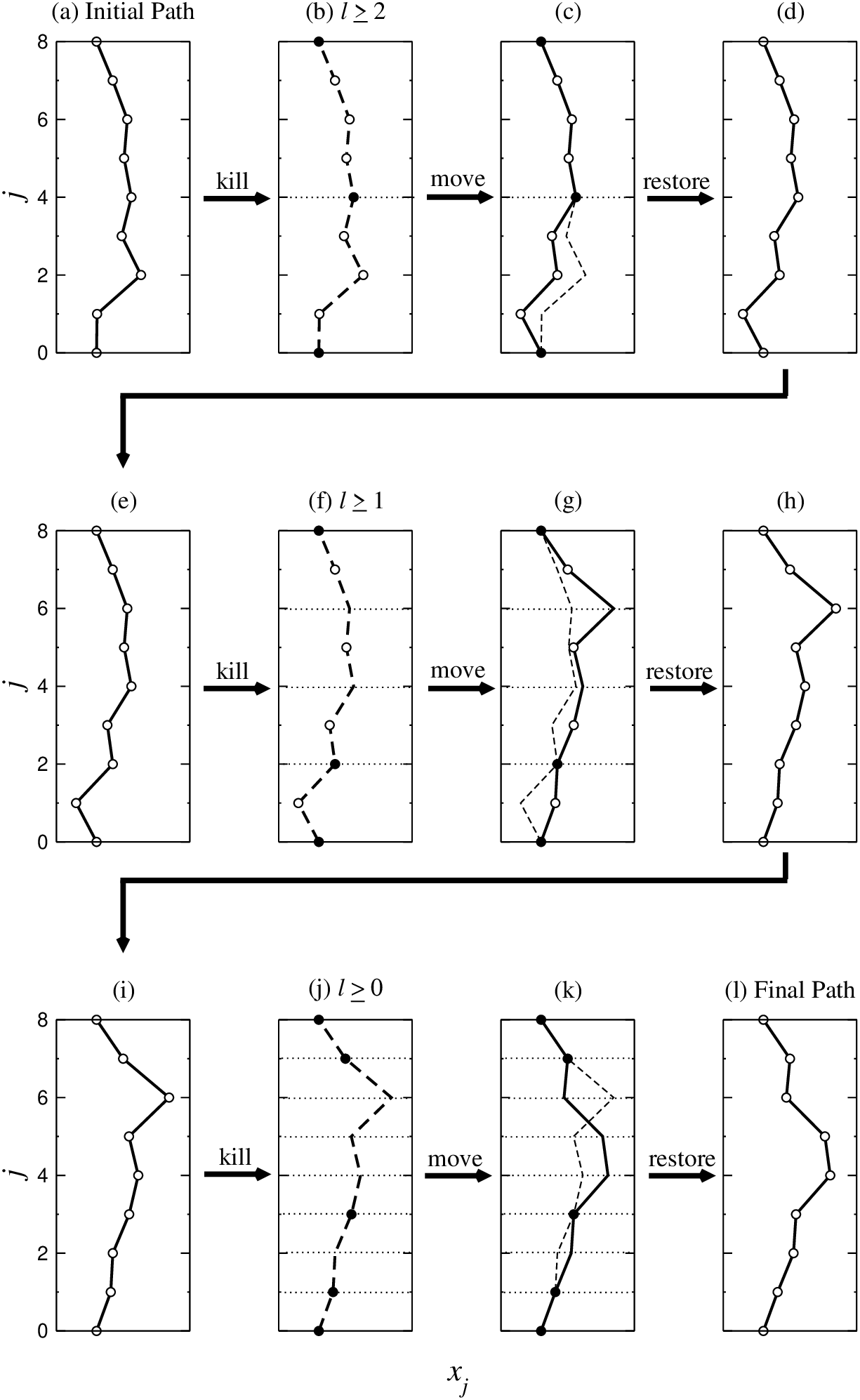}
\caption[]{
\label{fig:multigrid}
Illustration of the multigrid procedure.  (See text for details.)
}
\end{figure}

The multigrid method makes use of the stochastic blocking procedure in the 
unigrid method, but forces the path segments to break up on a predefined set 
of length scales.  The procedure is illustrated pictorially for a path with
$L=3$ in Fig.~\ref{fig:multigrid}.

Starting with the initial path in panel~(a), we kill all $V(x_j)$ on levels 
$\ell \geq 2$.  This results in a frozen bead at $j=4$.  The open circles 
in panel~(b) indicate $V(x_j)$ that are kept alive.  When the new path segments
are generated, they must be accepted or rejected based on a Metropolis 
criterion involving all the live beads.  Panel~(c) shows the new path 
after the move: the new segment between $j$ = $0$ and $4$ is accepted based
on the three live beads at $j$ = $1$--$3$, but the other new segment between 
$j$ = $4$ and $8$ is rejected based on the three live beads at 
$j$ = $5$--$7$; as a result, the new path (solid line) between 
$j$ = 4 and 8 coincides with 
and the old one (dashed line).  Panel~(d) shows the new path on this level 
after all $V(x_j)$ are restored.  Then we move on to the next finer level.  
On this level, we kill all beads on level $\ell \geq 1$, namely at $j$ = 
2, 4 and 6.  This results in the frozen (solid circles), deleted (no circles) 
and live (open circles) beads indicated in panel~(f).  Two new segments are
generated independently 
and accepted based on the live bonds at $j$ = 1 and $j$ = 3, 5 and 
7.  All beads are then restored and the procedure repeats on the finest level 
in the bottom row of Fig.~\ref{fig:multigrid}, starting with killing all 
$V(x_j)$ for $\ell \geq 0$.

Notice that the method presented here is very different from a 
method previously described by Janke and Sauer \cite{93jan499,96jan488}, 
which they also call a ``multigrid path integral method''.

\section{Results} \label{sect:results}

We have carried out numerical tests on our algorithm for 
several 1-dimensional quantum systems and compared them against other 
methods, including:
\begin{enumerate}
\item Metropolis: a conventional Metropolis algorithm based on 
single-bead moves in the $x$-coordinates; 
\item Bisection: the bisection algorithm as described by Ceperley 
in \cite{95cep279}; 
\item Unigrid: the unigrid algorithm as described above; and 
\item The so-called ``multigrid path integral method'' of Janke and Sauer 
\cite{93jan499,96jan488}.
\end{enumerate}

The tests were carried out on different model systems with symmetric 
single- and 
double-well potentials.  In the simulations, we have adopted a system of 
dimensionless units in which $m = \hbar = 1$.   For all tests, $\beta = 10$, 
yielding a thermal wavelength of approximately $1.6$  which sets the 
length scale of the quantum dispersion due to the kinetic part of the action.  
For all of the systems studied, the ground state dominates at this temperature.
Because the potentials are all symmetric, $\langle x \rangle$ should be 
$0$; therefore, how quickly the measured $\langle x \rangle$ goes to the
exact value of $0$ will provide a good estimate of the efficiency of each 
algorithm.  Alternatively, if we calculate the quantity $\bar x \equiv 
\sum_{j=1}^P x_j$ after each Monte Carlo step, we can determine the 
efficiency of the algorithm by examining how rapidly $\bar x$ 
fluctuates around the exact value $\langle x \rangle = 0$.

\subsection{Harmonic Potential (Model A): $V(x) = \frac{1}{2} x^2$}

\begin{figure}
\includegraphics[width=0.8\columnwidth]{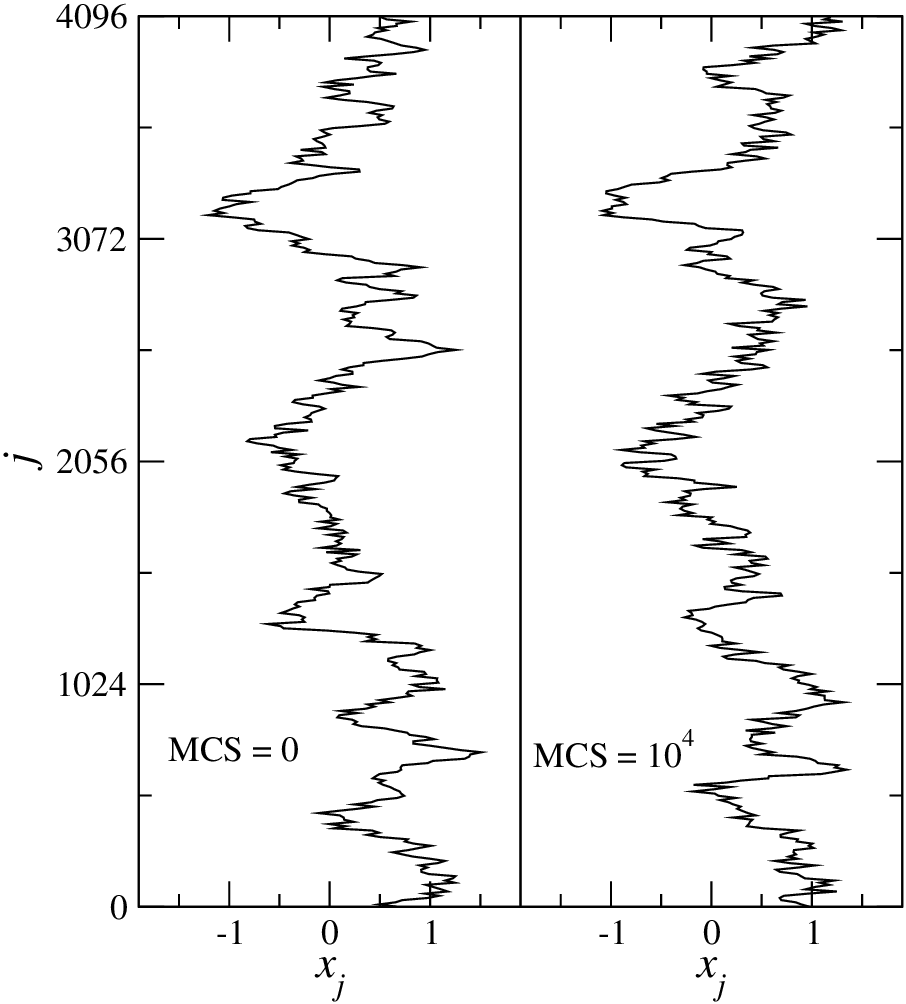}
\caption[]{
\label{fig:metropolis}
Initial and final paths after $10^4$ MCS in a conventional Metropolis
Monte Carlo simulation of Model~A
using a discretization of $L=12$.  The configuration has hardly 
moved, indicating the severity of the slowing-down problem.
}
\end{figure}

Model~A is a simple harmonic potential.   Figure~\ref{fig:metropolis} 
shows the initial and
final quantum paths after $10^4$ single-particle Metropolis Monte Carlo steps 
(one MCS is defined as having 
every bead on the path subjected to one trial move on the average).  
There is very little movement in the configuration of the whole path.  
Clearly, the conventional Metropolis algorithm is highly ineffective.  
$P = 4096$ beads, or $L = 12$, were used to represent the path here.
With $P$ of this magnitude, the bead-to-bead dispersion is merely 0.05, 
roughly 3\% of the thermal wavelength, making the harmonic bonds between
successive beads along the ring extremely stiff.

\begin{figure}
\includegraphics[width=0.8\columnwidth]{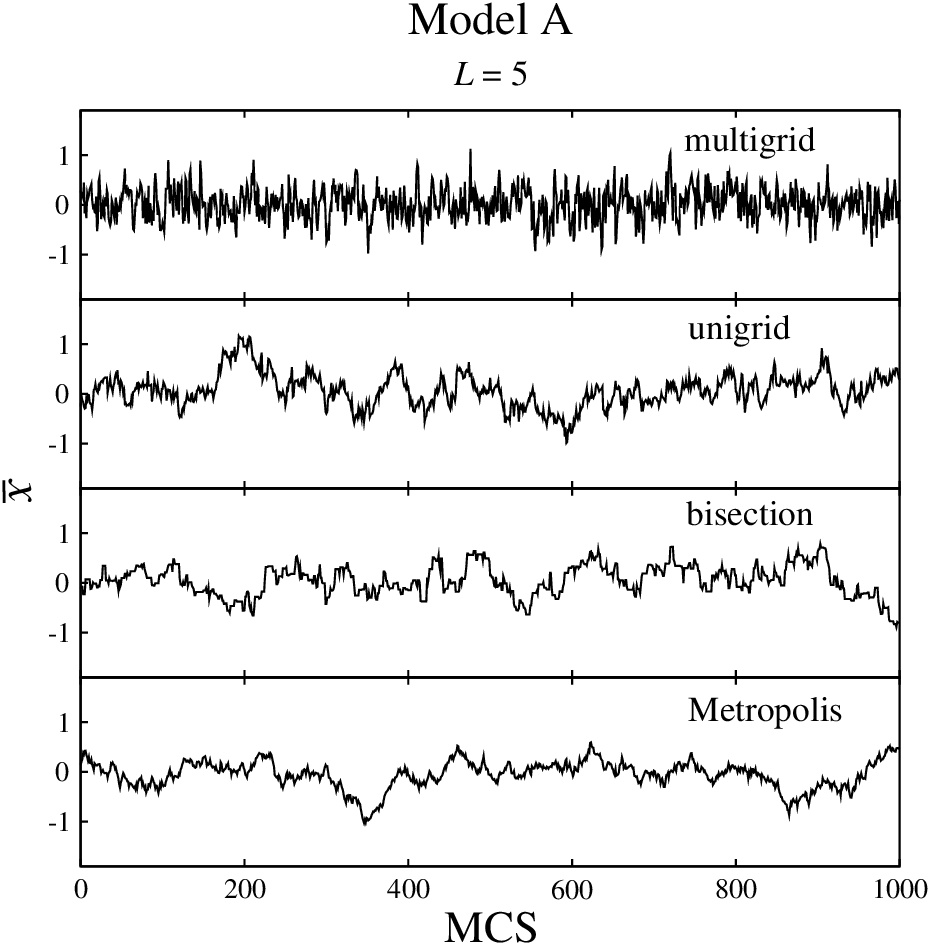}
\caption[]{
\label{fig:pot0}
Measurement of $\bar x$ after each MCS during the course
of the MC simulation for Model~A using the Metropolis, bisection, 
unigrid and multigrid methods.  The results should fluctuate around the 
exact answer $\langle x \rangle = 0$.  The high-frequency fluctuations
in the results from the multigrid method indicates its high efficiency. 
$L=5$ was used to discretize the path.
}
\end{figure}

\begin{figure}
\includegraphics[width=0.8\columnwidth]{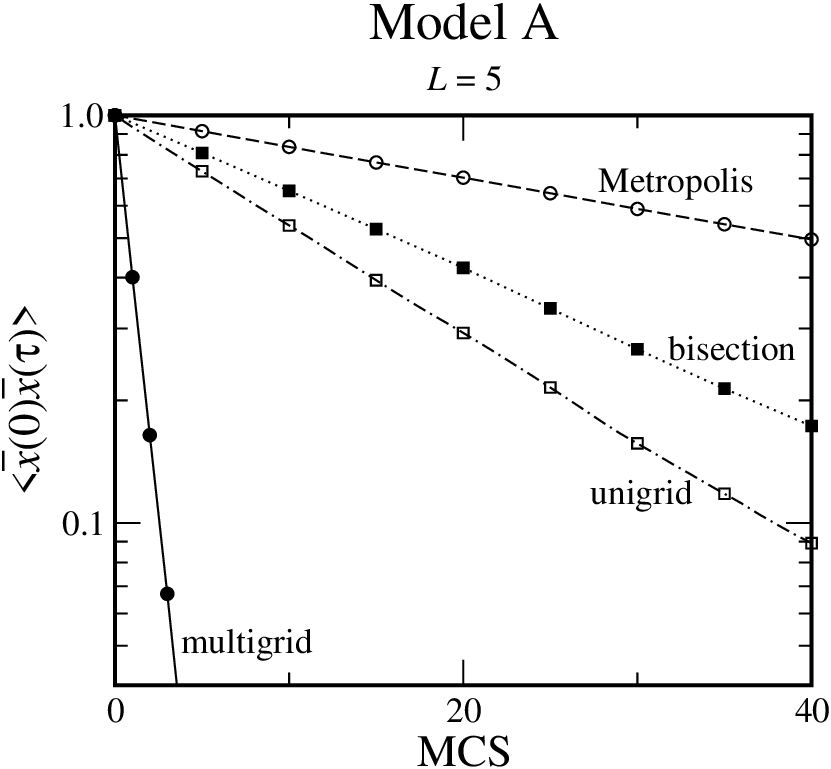}
\caption[]{
\label{fig:corr}
Autocorrelation functions of $\bar x$ for the data in Fig.~\ref{fig:pot0}.
The multigrid method has the fastest decay time.
}
\end{figure}

The discretization $L=12$ is much larger than what is needed for the path 
integral results to converge to the continuum limit for this model.  
The minimum required $L$ is 5.  Figure~\ref{fig:pot0} shows $\bar x$ 
as a function of MCS for the four methods for $L=5$.  Visually, we can see that
the multigrid method is the most efficient, the unigrid and the bisection 
methods are comparable and slower than the multigrid method, and the 
Metropolis method is the least efficient.  To get a more precise measure of
the efficiencies, the data in Fig.~\ref{fig:pot0} were autocorrelated and the 
correlation functions are shown in Fig.~\ref{fig:corr} for the four methods.  
The general conclusions we obtained from a visual inspection of 
Fig.~\ref{fig:pot0} are confirmed by Fig.~\ref{fig:corr} --
the multigrid method has the fastest decay time constant, approximately 1 MCS, 
and therefore is the most efficient.

\begin{figure}
\includegraphics[width=0.8\columnwidth]{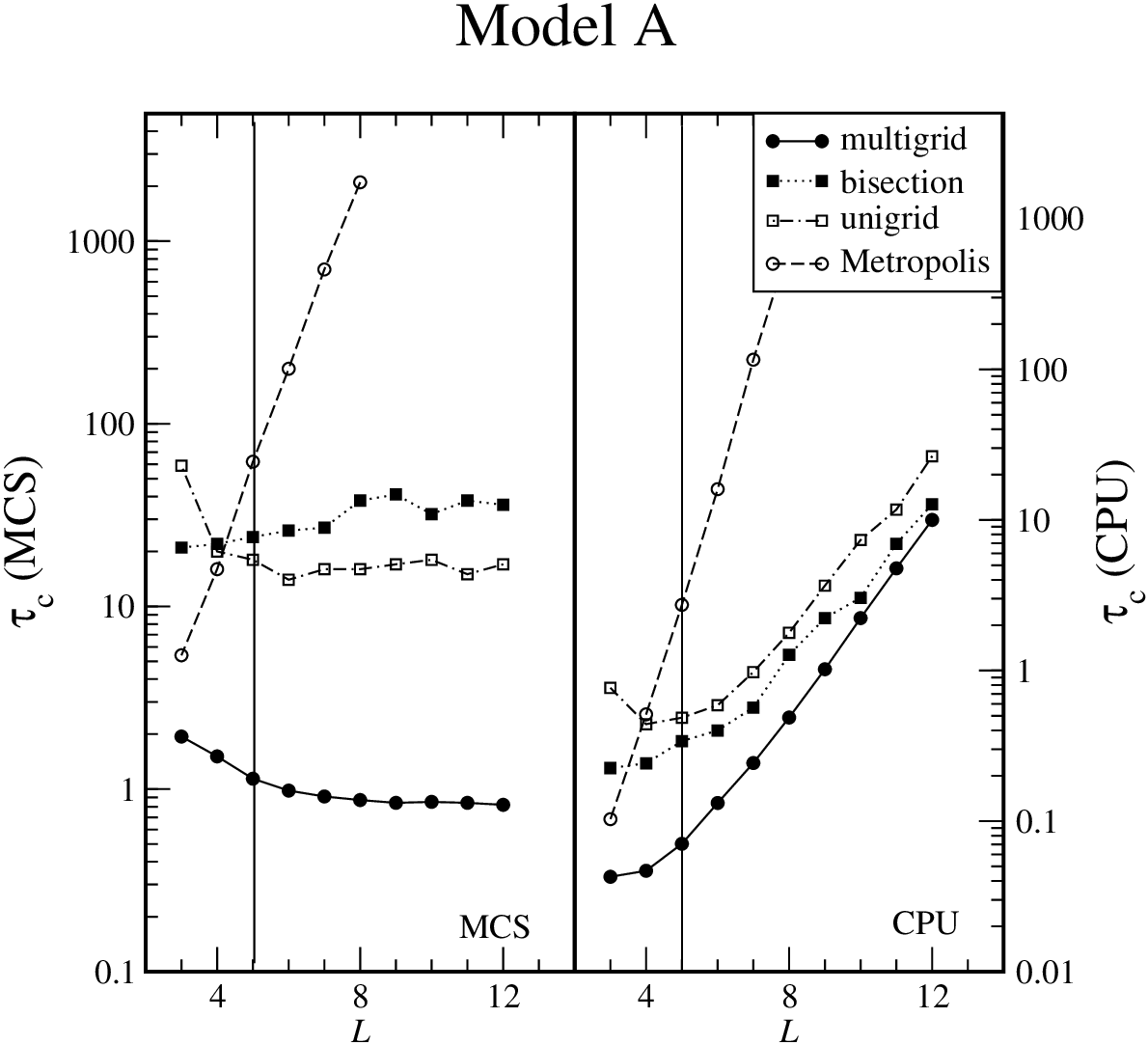}
\caption[]{
\label{fig:tau0}
Correlation times $\tau_c$ from four different MC methods at 
different discretizations $L$ for Model~A.  The vertical line at $L=5$ 
indicates the minimum $L$ needed for the path integral results to 
converge to the continuum limit.  The left panel shows correlation times in 
units of MCS.  The right panel shows the same but in units of 
actual CPU millisecond.  
(The segment length used for the bisection method was $2^{L-2}$ for each $L$.)
}
\end{figure}

The efficiencies of the four different methods for other values of $L$ are
shown in Fig.~\ref{fig:tau0}.  The left panel of Fig.~\ref{fig:tau0} shows 
the decay time constants $\tau_c$ in units of MCS 
from the autocorrelation function of $\bar x$ 
for the four methods at different path discretizations $L$.  The vertical 
line at $L=5$ indicates the minimum value of $L$ needed for the path integral 
results to converge for this model.  
Everything on the left of this line does not converge 
to the correct continuum limit.  Therefore, the only relevant results are 
those to the right of this line.

From a computational standpoint, the ultimate measure of real efficiency is 
the correlation time measured in 
{\em actual CPU cycles} (i.e. $\tau_c$ in units of MCS multiplied by CPU time 
per MCS) since difference algorithms 
will incur different CPU time per MCS.  The right panel in Fig.~\ref{fig:tau0}
shows correlation times in actual CPU milliseconds.  The real efficiency of the
multigrid method at the minimum required discretization for convergence 
($L=5$) is about a factor of 4 better than its nearest competitor, 
the bisection method, and a factor of 40 better than the standard 
Metropolis method.

This model has essentially one length scale dictated by the confinement 
effect of the potential.  For the particular parameters chosen for this
model, the confinement length scale of the potential turns out to be 
quite similar to the natural thermal wavelength of the free-particle 
path.  For this situation, a relatively small $L$ is sufficient for the
path integral to converge to its continuum limit.  Moreover, the four 
different methods, with their correlation times spanning only one and a half 
orders of magnitude, do not exhibit vastly different efficiencies.

These conclusions drawn from Model~A are certainly not universal.  The 
relative efficiencies of the various methods will depend on a number of 
factors, such as the length scale of the potential and the temperature.  
The additional 
examples presented below demonstrate this.  But in all the models considered, 
the multigrid method was always the most efficient by an order of magnitude 
or more compared to any other method.

\subsection{Compressed Harmonic Potential (Model B): 
$V(x) = \frac{1}{2} (3x)^2$}

\begin{figure}
\includegraphics[width=0.8\columnwidth]{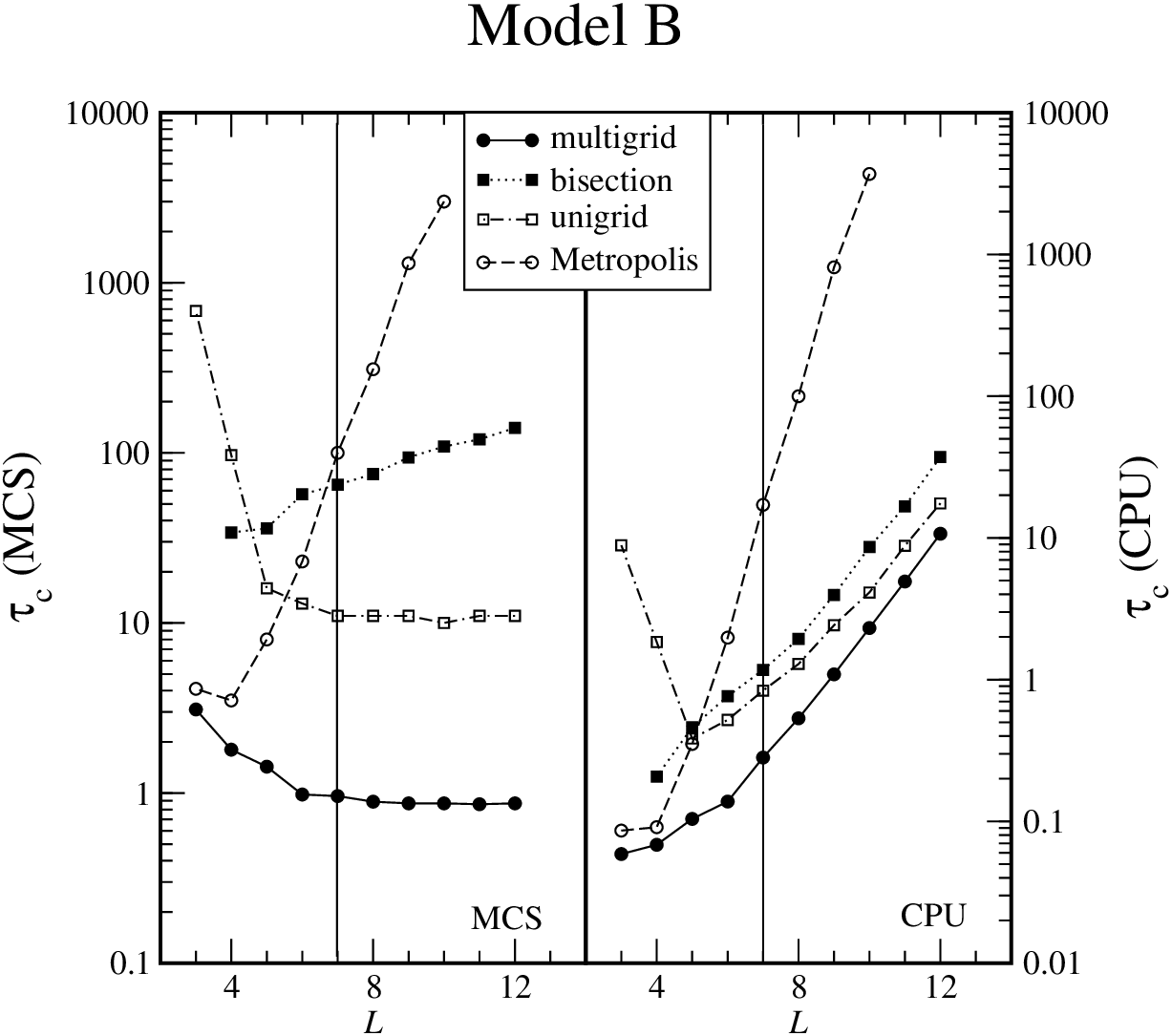}
\caption[]{
\label{fig:tau2}
Correlation times $\tau_c$ from four different MC methods at 
different discretizations $L$ for Model~B.  The minimum discretization 
required for convergence is $L=7$, indicated by the vertical line.  
Left and right panels show $\tau_c$ in units of MCS and CPU millisecond, 
respectively.
(The segment length used for the bisection method was $2^{L-3}$ for each $L$.)
}
\end{figure}

In Model~B, the length scale of the potential is substantially smaller than
the thermal wavelength.  This situation is quite typical in 
real condensed-phase quantum systems.  
The decay time constants of the autocorrelation functions of $\bar x$ are
summarized in Fig.~\ref{fig:tau2} for different $L$ and the four MC 
methods.  For this model, $L=7$ is the minimum required for convergent 
path integral results.  At or above this value of $L$, the multigrid method 
outperforms all the other methods in both MCS and CPU efficiencies.  
For this model, 
the second most efficient method is the unigrid method, which performs at
about a factor of 4 poorer than the multigrid method at $L=7$ in terms of 
real CPU efficiency.  On the other hand, the Metropolis method is a factor 
of 80 less efficient.

\subsection{Double-Well Potential (Model C): $V(x) = -3 x^2 + x^4$}

Model~C is a double-well potential with a moderate barrier.  The inter-well
separation is somewhat longer than the thermal wavelength but not by much.  
This is the first model that has at least two length scales due to 
the bistable nature of the potential.  Because of the moderate 
barrier, the length scales of the intra-well and inter-well quantum 
fluctuations are in reality not very different from each other.  

\begin{figure}
\includegraphics[width=0.8\columnwidth]{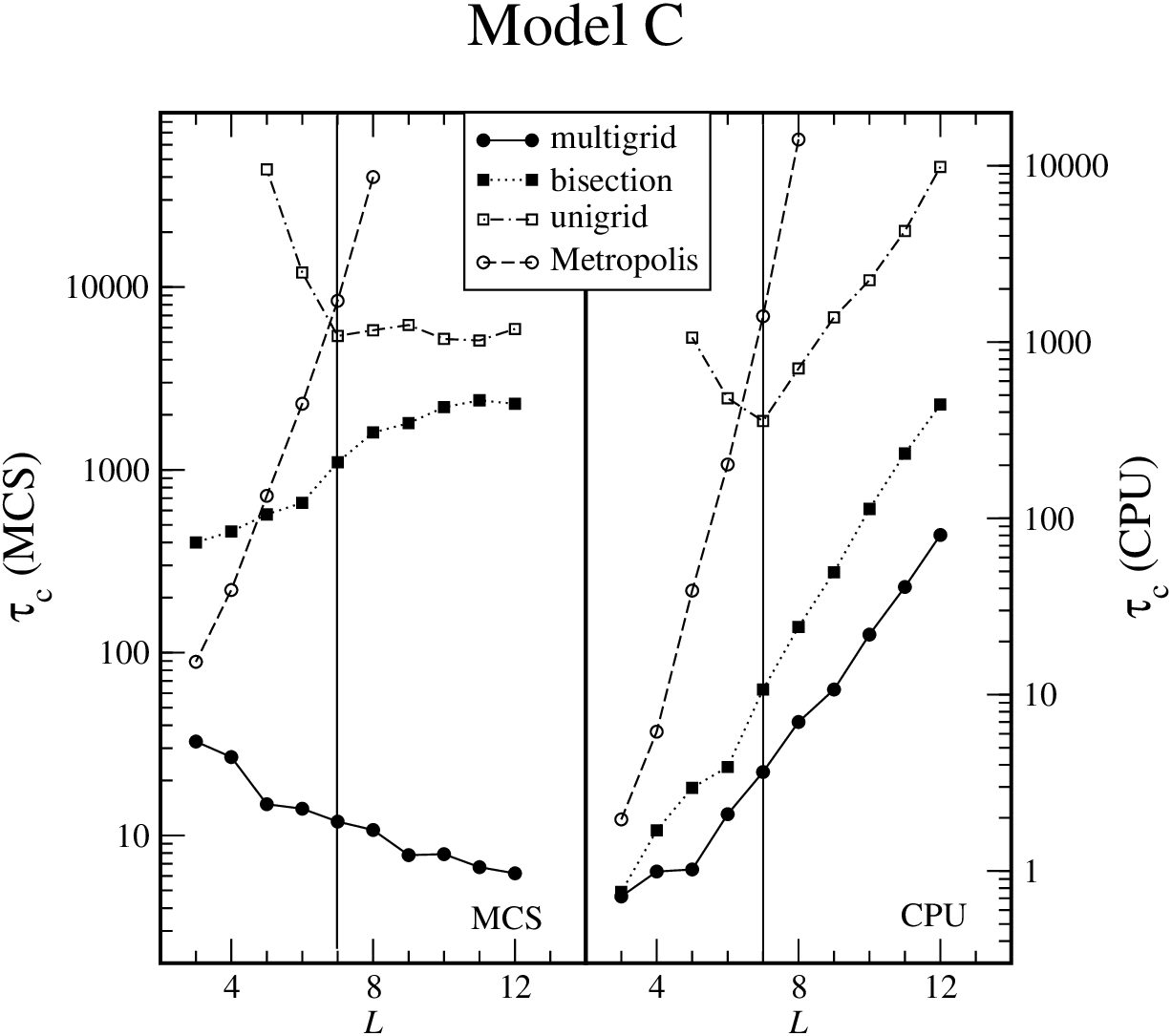}
\caption[]{
\label{fig:tau5}
Correlation times $\tau_c$ from four different MC methods at 
different discretizations $L$ for Model~C.  The minimum discretization 
required for convergence is $L=7$, indicated by the vertical line.  
Left and right panels show $\tau_c$ in units of MCS and CPU millisecond, 
respectively.
(The segment length used for the bisection method was $2^{L-2}$ for each $L$.)
}
\end{figure}

The decay time constants of the autocorrelation functions of $\bar x$ are
summarized in Fig.~\ref{fig:tau5} for different $L$ and the four MC 
methods.  For this model, $L=7$ is the minimum required for convergent 
path integral results.  At or above this value of $L$, the multigrid method 
outperforms all the other methods in both MCS and CPU efficiencies.  
For this model, 
the next most efficient method is the bisection method, which performs at
about a factor of 3 poorer than the multigrid method at $L=7$ in terms of 
real CPU efficiency.

\subsection{Compressed Double-Well Potential (Model D): 
$V(x) = -3 (2x)^2 + (2x)^4$}

Model~D has the same double-well potential as Model~C, but the potential 
is compressed in the $x$-direction.  In addition to having two distinct 
length scales, this model is further complicated by having a rather severe 
confinement effect because the length scale of the potential is much smaller 
than the natural thermal wavelength of the free-particle path.  The result 
is that the minimum discretization 
required for convergence becomes larger ($L=8$) and 
the multigrid method becomes increasingly advantageous 
compared to the other methods.
On the other hand, the bisection method suffers here, because it generates 
path segment of only one predefined length scale.  When that length scale 
is adjusted for maximum optimal efficiency, it would match one but 
misses all the other relevant length scales present in the problem 
\cite{footnote1}.  

\begin{figure}
\includegraphics[width=0.8\columnwidth]{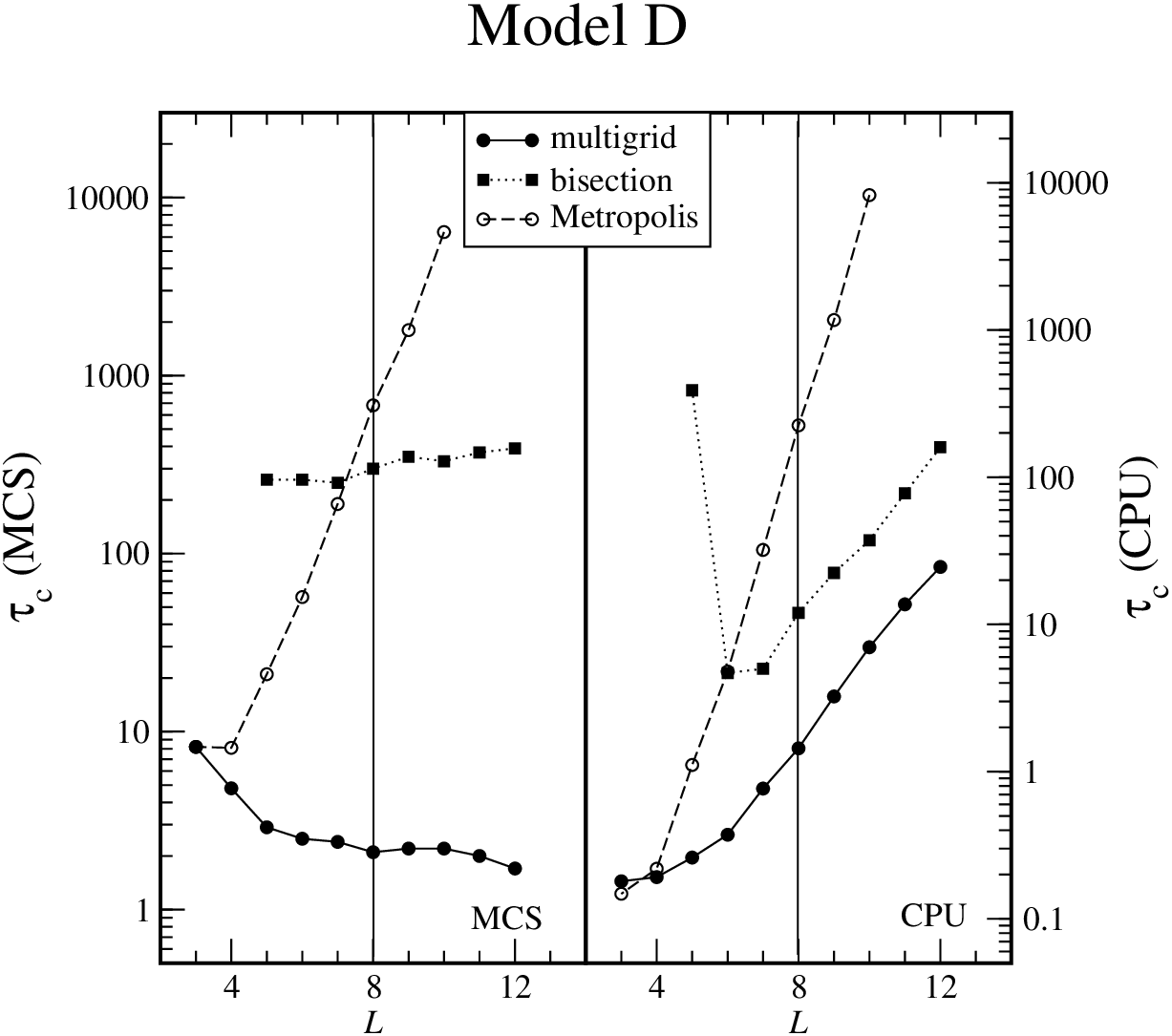}
\caption[]{
\label{fig:tau6}
Correlation times $\tau_c$ from three different MC methods at 
different discretizations $L$ for Model~D.  The minimum discretization 
required for convergence is $L=8$, indicated by the vertical line.  
Left and right panels show $\tau_c$ in units of MCS and CPU millisecond, 
respectively.
(The segment length used for the bisection method was $2^{L-4}$ for each $L$.)
}
\end{figure}

The decay time constants of the autocorrelation functions of $\bar x$ are
summarized in Fig.~\ref{fig:tau6} for different $L$ and three MC 
methods.  The results from the unigrid method are not shown 
because for every $L$ the decay time in the unigrid results is greater than 
$10^4$ MCS and we were unable to equilibrate the unigrid simulations.  
The multigrid method 
outperforms the other two methods in both MCS and CPU efficiencies.  
For this model, 
the next most efficient method is the bisection method, which performs at
about a factor of 10 poorer than the multigrid method at $L=8$ in terms of 
real CPU efficiency.

\subsection{Model of Janke and Sauer (Model E): $V(x) = -0.5 x^2 + 0.04 x^4$}

\begin{figure}
\includegraphics[width=0.8\columnwidth]{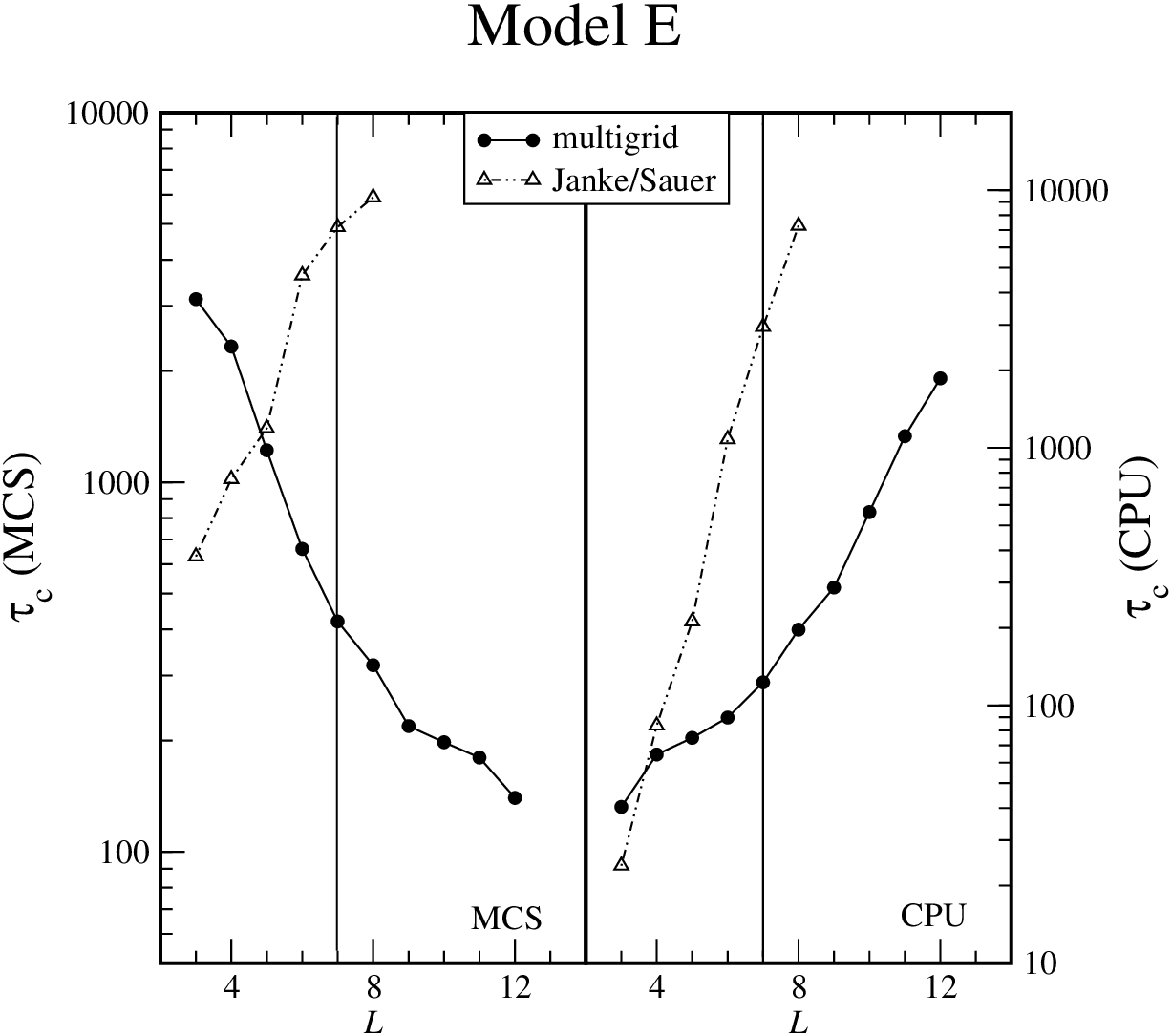}
\caption[]{
\label{fig:tau9}
Correlation times $\tau_c$ from three different MC methods at 
different discretizations $L$ for Model~E.  The minimum discretization 
required for convergence is $L=7$, indicated by the vertical line.  
Left and right panels show $\tau_c$ in units of MCS and CPU millisecond, 
respectively.
}
\end{figure}

To make contact with the results of Janke and Sauer 
\cite{93jan499,96jan488}, we carried out simulations on the double-well
potential studied in their papers, using the same parameters they have used.  
The decay time constants of the autocorrelation functions of $\bar x$ are
summarized in Fig.~\ref{fig:tau9} for different $L$.  
$L=7$ is the minimum required for convergent path integral results.   
The Metropolis, unigrid and bisection results are not shown because 
their decay times are all greater than $10^4$ MCS for all values of $L$.  
At $L=7$, the 
multigrid method is about an order of magnitude more efficient than the
Janke/Sauer method.  But notice that the CPU correlation time for this model 
is much larger ($\tau_c \approx 100$ CPU ms) compared to all the previous 
models.  Therefore, this model potential seems to present a 
slightly more challenging problem even 
for the multigrid method.

\subsection{An Electron in the Field of Two Positive Ions (Model F)}

\begin{figure}
\includegraphics[width=0.8\columnwidth]{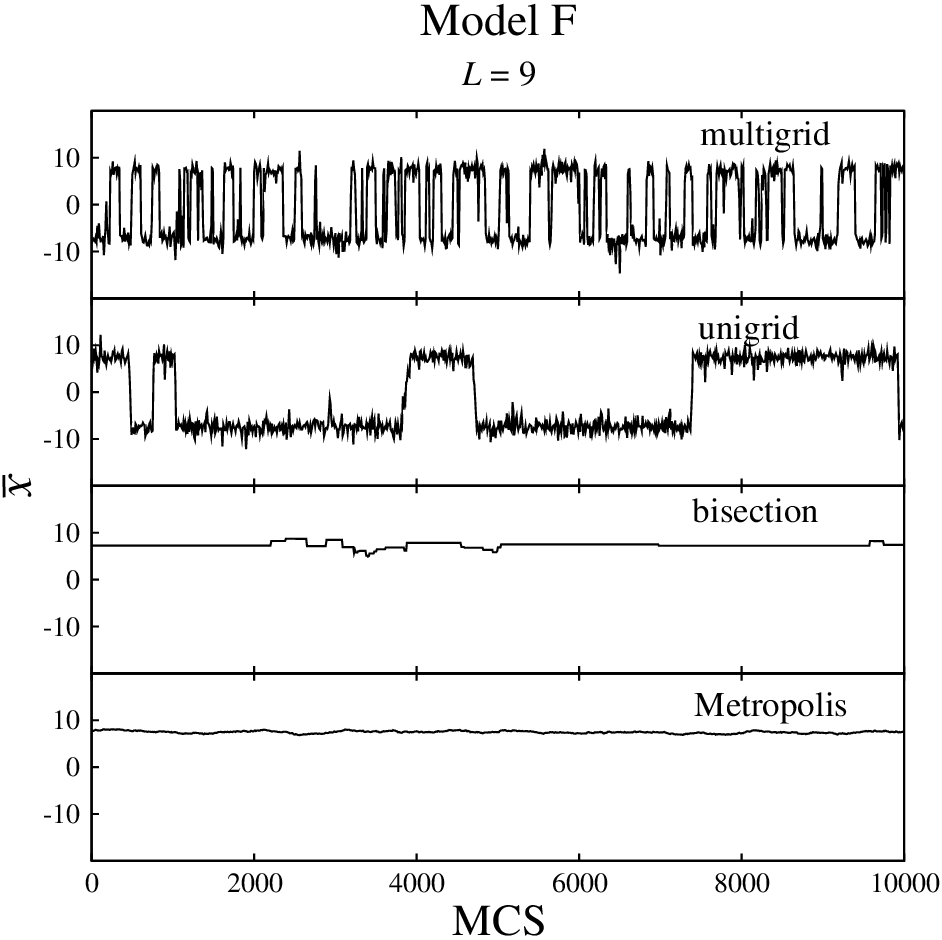}
\caption[]{
\label{fig:pot10}
Measurement of $\bar x$ after each MCS during the course
of the MC simulation for Model~F using the Metropolis, the bisection, 
the unigrid and the multigrid methods using a discretization of $L=9$.
}
\end{figure}

\begin{figure}
\includegraphics[width=0.8\columnwidth]{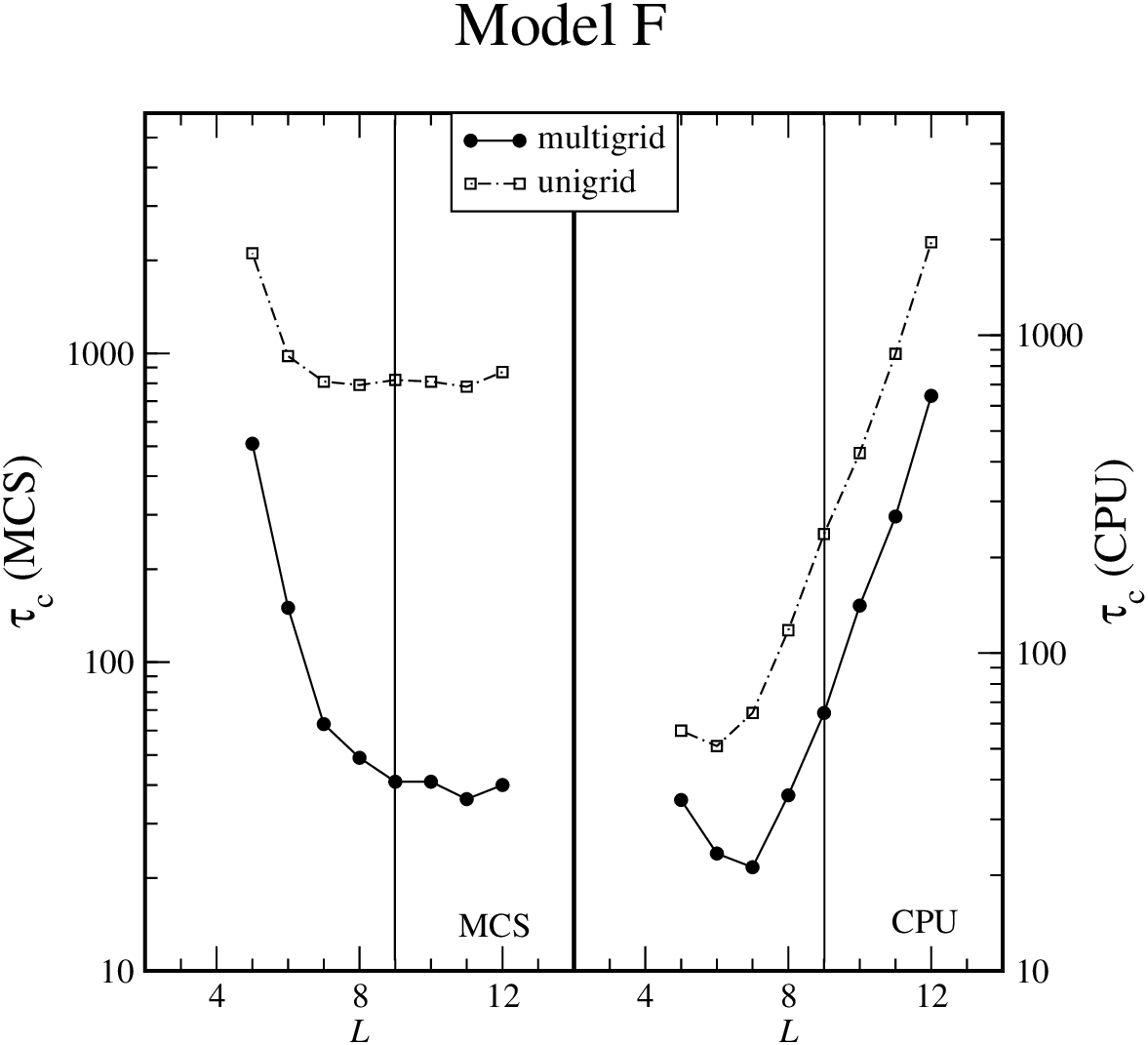}
\caption[]{
\label{fig:tau10}
Correlation times $\tau_c$ from the multigrid and the unigrid methods 
for Model~F.  The minimum discretization 
required for convergence is $L=9$, indicated by the vertical line.  
Left and right panels show $\tau_c$ in units of MCS and CPU millisecond, 
respectively.
}
\end{figure}

This last model consists of a 1-dimensional electron in the field of two 
ions of $+2$ charge separated by a distance of 15~\AA.  
To prevent the electron path from collapsing onto either ion, we 
set up a 0.25~\AA\  hard core radius around each.  
The temperature is 300~K and the dielectric constant is 78.
Similar to some of the other models already considered, this model consists
of a bistable potential with a high barrier between the two wells.  
This model is however qualitative quite different from the others, 
because the coulombic
potential centered on each ion is {\em much} narrower in comparison with the
fairly open well bottoms in the other models.  This results in a strong 
confinement effect on the electron path.  The natural quantum 
fluctuations of the path are of a longer scale than the width of 
the potential wells.  
Figure~\ref{fig:pot10} shows $\bar x$ as a function of MCS for the four
methods at $L=9$, the minimum required discretization needed for convergence.  
Both the bisection and the Metropolis methods failed to equilibrium in  
10000 MCS.  The decay time constants of the autocorrelation function 
of $\bar x$ are shown in Fig.~\ref{fig:tau10} for the multigrid and 
the unigrid methods.

\section{Conclusion}

We have described a new Monte Carlo method for simulating imaginary-time
path integrals.  The new method uses a combination of a stochastic
blocking algorithm and multigrid ideas to completely eliminate the
slowing-down problem in the sampling of discretized quantum paths
near the continuum limit.  The method has been tested on several 
1-dimensional quantum systems and found to exhibit highly ergodic
dynamics.  The new method offers distinct advantages over other
methods in cases where the length scale of the potential is
smaller than the quantum dispersion of the path, a situation that 
is quite typical in real condensed-phase quantum systems.  On the 
other hand, for systems with bistable potentials with length 
scales much larger than the quantum dispersion of the path, the new method, 
though better than all the other methods, has only limited utility.

\section{Acknowledgments}

This work was supported by the National Science Foundation (CHE-9970766). 



\begin{thebibliography}{99}
\bibitem{footnote1} 
It may be possible that the efficiency of the bisection method can be 
improved by incorporating a multigrid element into it.
\end{thebibliography}
\end{document}